\documentclass[structabstract]{aa}  
\usepackage{graphicx}
\usepackage[]{natbib}
\usepackage{txfonts}
\usepackage[normalem]{ulem}
\begin{document}
   \title{On the origin of the jet-like radio/X-ray morphology of G290.1--0.8}

   \author{F. Garc\'{\i}a \inst{1,2,}\thanks{Fellow of CONICET, Argentina.}, J.A. Combi\inst{1,2}, J.F. Albacete-Colombo\inst{3}, G.E. Romero\inst{1,2}, F. Bocchino\inst{4}, J. L\'opez-Santiago\inst{5}
  }

\authorrunning{Garc\'{\i}a et al.}

\titlerunning{On the origin of the jet-like radio/X-ray morphology of G290.1--0.8} 

\offprints{F. Garc\'{\i}a}  

\institute{Instituto Argentino de Radioastronom\'{\i}a (CCT La Plata, CONICET), C.C.5, (1894) Villa Elisa, Buenos Aires, Argentina.\\
\email{[fgarcia,jcombi,romero]@iar-conicet.gov.ar}
\and
Facultad de Ciencias Astron\'omicas y Geof\'{\i}sicas, Universidad Nacional de La Plata, Paseo del Bosque, B1900FWA La Plata, Argentina.
\and
Centro Universitario Regional Zona Atl\'antica (CURZA). Universidad Nacional del COMAHUE, Monse\~nor Esandi y Ayacucho (8500), 
Viedma (Rio Negro), Argentina.\\
\email{donfaca@gmail.com}
\and
INAF-Osservatorio Astronomico di Palermo, Piazza del Parlamento 1, 90134, Palermo, Italy.\\
\email{bocchino@astropa.inaf.it}
\and
Departamento de Astrof\'{\i}sica y Ciencias de la Atm\'osfera, Universidad Complutense de Madrid, E-28040, Madrid, Spain.\\
\email{jls@astrax.fis.ucm.es}     
             }

   \date{Received 3 February 2012 ; accepted 11 September 2012}

% \abstract{}{}{}{} 
% 5 {} token are mandatory
 
  \abstract
	{The origin and evolution of supernova remnants of the mixed-morphology class is not well understood. Several remnants present distorted radio or X-ray shells with jet-like structures. G290.1$-$0.8 (\object{MSH 11-61}A) belongs to this particular class.}
	{We aim to investigate the nature of this supernova remnant in order to unveil the origin of its particular morphology. We based our work on the study of the X-ray emitting plasma properties and the conditions imposed by the cold interstellar medium where the remnant expanded.} 
   {We use archival radio, \ion{H}{i} line data and X-ray observations from \textit{XMM-Newton} and \textit{Chandra} observatories, to study G290.1$-$0.8 and its surrounding medium in detail. Spatially resolved spectral analysis and mean photon energy maps are used to obtain physical and geometrical parameters of the source. Radio continuum and \ion{H}{i} line maps give crucial information to understand the radio/X-ray morphology.}
	{The X-ray images show that the supernova remnant presents two opposite symmetric bright spots on a symmetry axis running towards the North West-South East direction. Spectral analysis and mean photon energy maps confirm that the physical conditions of the emitting plasma are not homogeneous throughout the remnant. In fact, both bright spots have higher temperatures than the rest of the plasma and its constituents have not reached ionization equilibrium yet. \ion{H}{i} line data reveal low density tube-like structures aligned along the same direction.
	This evidence supports the idea that the particular X-ray morphology observed is a direct consequence of the structure of the interstellar medium where the remnant evolved. However, the possibility that an undetected point-like object, as a neutron star, exists within the remnant and contributes to the X-ray emission cannot be discarded. Finally, we suggest that a supernova explosion due to the collapse of a high-mass star with a strong bipolar wind can explain the supernova remnant morphology.}
{}

%   \maketitle
	\maketitle
%________________________________________________________________

\section{Introduction}

%%%%%%%%%%%%%%%%%%%%%%%%%%%%%%%%%%%%%%%%%%%%%%%%%%%%%%%
\begin{table*}
\caption{Details of the {\it Chandra} and {\it XMM-Newton} observations of G290.1$-$0.8.}
\label{obs}\centering
\begin{center}
\begin{tabular}{l c c l c c c c}
\hline\hline
Satellite& \multicolumn{2}{c}{{\it Chandra}}&& \multicolumn{2}{c}{{\it XMM-Newton}} \\ 
\cline{2-3} \cline{5-6}
Obs-Id     & 2754 & 3720 && 0111210201 & 0152570101  \\ 
Date	 & 16/07/2002  & 21/07/2002	&& 28/07/2002 & 21/07/2002\\
Start Time	 [UTC]	& 03:12:39   & 01:45:38 && 02:29:26   & 16:56:39	\\
Camera    	 & ACIS-235678 & ACIS-235678	&& pn/MOS1,2& MOS1,2		\\
Filter 		 &  $--$	       	&	$--$			&& MEDIUM	& MEDIUM 	\\
Modes (read/data) & TIMED/VFAINT & TIMED/VFAINT && PFWE	& PFWE		\\
Offset 		 &  ccd7 - on axis   & ccd7 -on axis && on-axis& on-axis 	\\
Exposure [ks] & 29.71   & 	33.66	&& 6.04-10.8-10.8   & 62.2-62.3		\\
GTI	[ks]	& 26.04  & 30.59	&& 5.89-10.6-10.6 &  60.6-60.6		\\
\hline
\end{tabular}
\end{center}
\tablefoot{All observations were taken from the respective satellite databases. PFWE refers to the Prime Full Window Extended observation mode. Pointing of {\it Chandra} observations are 
$\alpha$= 11$^h$02$^m$57$\fs$20, $\delta$=$-$60$\degr$54$\arcmin$34$\farcs$0 (J2000.0).}
\label{obstable}
\end{table*}
%%%%%%%%%%%%%%%%%%%%%%%%%%%%%%%%%%%%%%%%%%%%%%%%%%%%%%%%%

%%%%%%%%%%%%%%fig-1%%%%%%%%%%%%%%%
\begin{figure*}
\centering
\includegraphics[width=18.3cm]{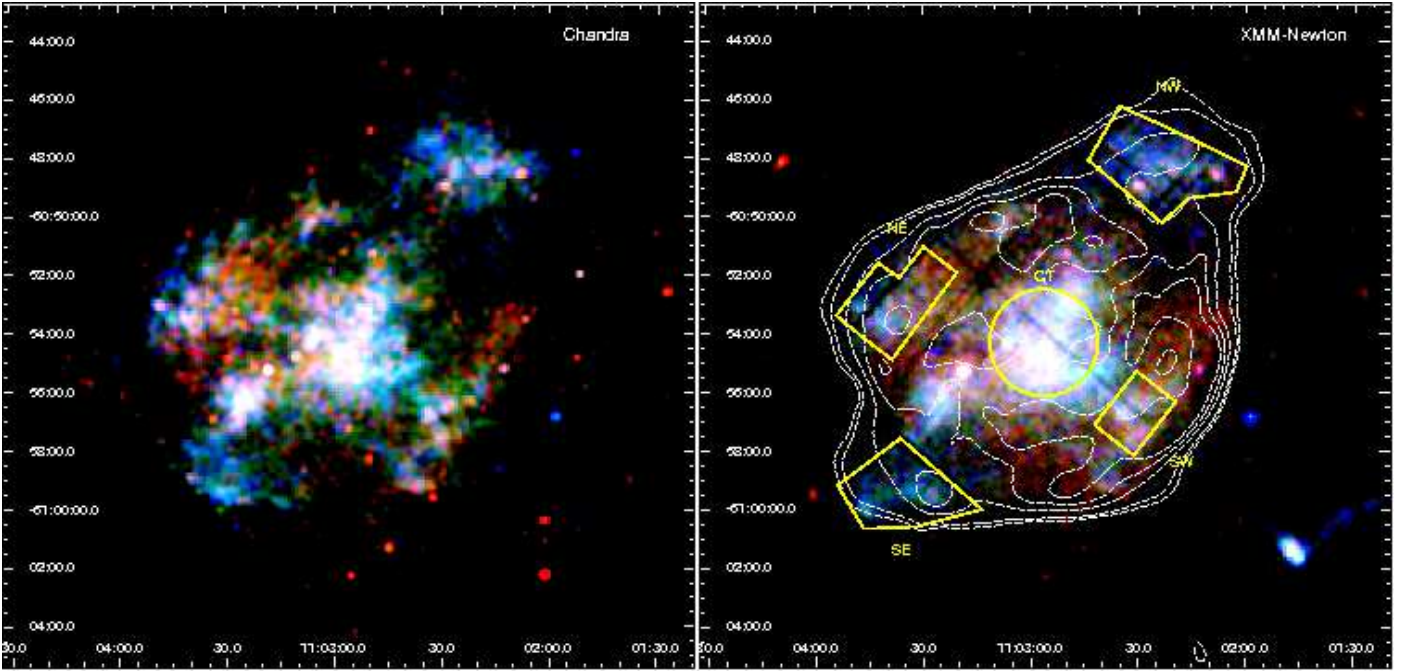}
   \caption{{\it Chandra} (left panel) and {\it XMM-Newton} (right panel) images of G290.1$-$0.8 in the three X-ray energy bands: soft (0.5$-$1.2 keV) in red, medium (1.2$-$1.8 keV) in green, and hard (1.8$-$2.8 keV) in blue. Right panel also shows the radio contours at 843 MHz \citep{whiteoak1996} overlaid in white. The five regions selected for the extraction of photons for our spectral study are also indicated in yellow. In both images North is up and East is to the left.}
\label{fig1}
\end{figure*}
%%%%%%%%%%%%%%%%%%%%%%%%%%%%%%%%%%%%%%%%%%%

The origin of supernova remnants (SNRs) displaying a distorted radio or X-ray shell with two opposite, symmetric bright hot spots is not well understood. Possible physical scenarios to explain the observed morphology include: the SNR expansion into multiple cavities in the interstellar medium (ISM) \citep{braun1986,pineault1987,milne1989,dubner1994}, expansion of the SNR through the progenitor wind with axisymmetric density distribution \citep{blondin1996}, or the presence of a central source which produces collimated outflows that impact on the shell in two opposite directions 
\citep{murata1996}.

In the Galaxy, there are several SNRs with these morphological characteristics \citep{gaensler1998}.
G290.1$-$0.8 is a member of this peculiar class of sources. The object was first identified as a SNR on the basis of its non-thermal radio spectrum \citep{kesteven1968}. Through radio observations at 408 MHz \citep{kesteven1987} and 5 and 8 GHz \citep{milne1989}, it was classified as a shell-type SNR, with an angular size of $15\times10$~arcmin and a spectral index of $\alpha = -0.4$ ($S \propto \nu^{\alpha}$). Using high resolution radio continuum and CO data, \cite{filipovic2005} studied its morphology and kinematics in detail. They found that the SNR is associated with a dense molecular cloud
located to the South West. More recently, \cite{reynoso2006}, using \ion{H}{i} line and 20~cm radio-continuum observations performed with the Australia Telescope Compact Array (ATCA),  showed that the gas distribution and kinematics in front of the SNR are complex and estimated that it probably lies in the Carina arm, at a distance of 7$\pm$1~kpc. 

At X-ray energies, G290.1$-$0.8 was first studied by \cite{seward1990} with the {\it Einstein} Observatory. These observations permited to establish the center-filled nature of the X-ray emission. Later, \cite{slane2002} analyzed Advanced Satellite for Cosmology and Astrophysics ({\it ASCA}) observations of the object and found that the X-ray emission is of thermal nature. They also classified the remnant as a member of the mixed-morphology (MM) class \citep[see e.g.][]{rho1998}, and estimated that its age should be 10$-$20~kyr.

The question about the origin and evolution of G290.1$-$0.8 and its particular morphology is still unanswered. In this paper, we report the results of the analysis of {\it XMM-Newton} and {\it Chandra} observations of G290.1$-$0.8. We study the physical characteristics of the object and present a possible scenario to explain the origin of its elongated and axisymmetric morphology observed at radio and X-ray frequencies. The structure of the paper is as follows: in Sect.~2, we describe the {\it XMM-Newton} and {\it Chandra} observations and the data reduction process. In Sect.~3, we present the results of our radio/X-ray data analysis, including X-ray images, spectra, mean photon energy map and the study of the \ion{H}{i} line in the medium where the SNR is immersed. In Sect. 4, we discuss a possible scenario to explain the origin of the observed morphology and finally, we summarize our main conclusions in Sect.~5.

\section{Observations and data reduction}

The field of G290.1$-$0.8 has been observed by the {\it XMM-Newton} observatory twice. The observations were performed with the European Photon Imaging Camera (EPIC) that consists of three detectors (two MOS cameras \citep{turner2001} and one pn \citep{struder2001} camera) operating in the 0.2$-$15~keV energy range. Both observations have similar pointing coordinates ($\alpha$= 11$^h$03$^m$00$\fs$0, $\delta$=$-$60$\degr$54$\arcmin$00$\farcs$0 ; J2000.0), with the SNR placed at the center of the CCDs. Two {\it Chandra} X-ray observations with the Advanced CCD Image Spectrometer (ACIS-I) camera are also available. ACIS-I operates in the 0.1$-$10~keV energy band with high spatial resolution ($\sim$0.5~arcsec).
Such a large dataset give us the possibility, by the first time, to perform a detailed X-ray analysis of SNR G290.1$-$0.8. {\it XMM-Newton} data were analyzed with the {\it XMM-Newton} Science Analysis System (SAS) version 11.0.0 and the latest calibrations. {\it Chandra} observations were calibrated using CIAO (version 4.1.2) and CALDB (version 3.2.2). To exclude intense background flares periods, which could eventually affect the observations, we extracted light curves of photons above 10~keV from the entire field-of-view of the cameras, and excluded intervals up to $3\sigma$ to produce a Good Time Interval (GTI) file. Detailed information of the observations and the instrumental characteristics are given in Table~\ref{obstable}.

\section{Results}
\subsection {X-ray images}

The high spatial resolution and sensitivity of the X-ray observations allowed us to examine the X-ray morphology of G290.1$-$0.8 in depth. In Figure~\ref{fig1}, we show {\it Chandra} (left panel) and {\it XMM-Newton} (right panel) narrow-band images generated in the soft (0.5$-$1.2~keV), medium (1.2$-$1.8~keV) and hard (1.8$-$2.8~keV) energy bands. On the right panel of Figure~\ref{fig1}, we also present the 843 MHz radio contours overlaid (in white color). These radio contours correspond to the Supernova Remnant Catalog \citep{whiteoak1996} from the Molonglo Observatory Synthesis Telescope (MOST). To produce the X-ray image, the data of both MOS cameras were merged, exposure and vignetting corrected, and smoothed with a 3 pixel radius Gaussian filter.

The {\it XMM-Newton} and {\it Chandra} images reveal details of the X-ray structures of G290.1$-$0.8 that have not been found in previous X-ray studies \citep{slane2002}. The global shape of the SNR is elongated along its North West/South East axis with two opposite, symmetric bright spots almost dettached from the diffuse central component region. In the North East/South West direction there are also two ear-like structures diametrically opposed with respect to the geometric center of the SNR. The extended X-ray emission fills the entire interior of the outer radio boundary, being quite prominent at medium and hard energies. The southern, northern and external parts of the ear-like components are brighter in medium and hard energies relative to the total X-ray emission. Above $2.8$~keV, no emission from the SNR is detected in any of the X-ray observations. It is also remarkable that the presence of a compact point-like source (a putative pulsar) within the remnant could not be revealed in any of the X-ray images obtained from the available data.

\subsection{X-ray spectral analysis}

%%%%%%%%%%%%%table-2%%%%%%%%%
\begin{table*}
\caption{Spectral parameters of the diffuse X-ray emission of the five selected regions.}
\label{spec}
\begin{center}
\begin{tabular}{l | cccccc}
\hline\hline
Model \& Parameters & CT region & SW region & NE region & NW region & SE region \\
\hline
{\bf PHABS*VPSHOCK} &&&&&& \\
%\hline
$N_\mathrm{H}$ [10$^{22}$~cm$^{-2}$]&	 
0.62$\pm$0.01 &
0.61$\pm$0.03 &
0.50$\pm$0.02 &
0.43$\pm$0.02 &
0.61$\pm$0.04 & \\
kT [keV] &				 
0.593$\pm$0.008   &
0.57$\pm$0.02   &
0.63$\pm$0.02   &
0.89$\pm$0.03   &
0.84$\pm$0.05   & \\
O [O$_\odot$]  &
0.61$\pm$0.06  &  
0.9$\pm$0.2  &
0.4$\pm$0.1  &  
0.13$\pm$0.07 &  
0.3$\pm$0.1 &  \\
Ne [Ne$_\odot$]  &
0.26$\pm$0.03  &  
0.33$\pm$0.08  &
0.17$\pm$0.06  &  
0.13$\pm$0.08 &  
0.07$\pm$0.05 &  \\
Mg [Mg$_\odot$]  &
0.95$\pm$0.03  &  
0.81$\pm$0.07  &
0.68$\pm$0.05  &  
0.60$\pm$0.06  &  
0.61$\pm$0.06 &  \\
Si [Si$_\odot$]  &
1.76$\pm$0.04 &  
1.5$\pm$0.1  &
1.18$\pm$0.07  &  
1.15$\pm$0.06  &  
1.35$\pm$0.09 &  \\
S [S$_\odot$]  &
1.4$\pm$0.1 &  
1.3$\pm$0.3  &
1.13$\pm$0.2  &  
1.0$\pm$0.1 &  
1.0$\pm$0.2 &  \\
Fe [Fe$_\odot$]  &
0.080$\pm$0.005 &  
0.12$\pm$0.01  &
0.09$\pm$0.01  &  
0.02$\pm$0.01  &  
0.02$\pm$0.01 &  \\
$\tau$[$10^{13}$~s~cm$^{-3}$] &
\textgreater 5 &
\textgreater 5 &
4.15$\pm$0.5 &
0.12$\pm$0.08 &
0.03$\pm$0.01 &\\
Norm [$10^{-3}$] & 
13.2$\pm$0.4& 
2.3$\pm$0.2 & 
4.4$\pm$0.2 & 
3.0$\pm$0.2 & 
2.0$\pm$0.3 & \\
\hline
$\chi^{2}_{\nu}$ / d.o.f. &
1.58 / 345 &
1.13 / 337  &
1.00 / 343 &
1.32 / 342 &
1.12 / 340 &\\
\hline
Flux(0.3$-$1.0~keV)[10$^{-13}$~erg~cm$^{-2}$~s$^{-1}$] & 
84.5$\pm$0.5&
18.0$\pm$0.4 &
25.5$\pm$0.4 &
17.2$\pm$0.3  &
24.1$\pm$0.5 &\\
Flux(1.0$-$2.0~keV)[10$^{-13}$~erg~cm$^{-2}$~s$^{-1}$] & 
35.9$\pm$0.2&
6.18$\pm$0.2 &
10.3$\pm$0.2 &
9.07$\pm$0.2 &
6.70$\pm$0.2  &\\
Flux(2.0$-$3.0~keV)[10$^{-13}$~erg~cm$^{-2}$~s$^{-1}$] & 
5.28$\pm$0.04  &
0.80$\pm$0.02 &
1.71$\pm$0.03 &
2.59$\pm$0.05 &
1.42$\pm$0.03 &\\
\hline
Total Flux(0.3$-$3.0~keV)[10$^{-13}$~erg~cm$^{-2}$~s$^{-1}$] &
126$\pm$1 &
25.0$\pm$0.6 &
37.5$\pm$0.6 &
28.8$\pm$0.5 &
32.2$\pm$0.7 &\\
\hline
\end{tabular}
\label{spectable}
\tablefoot{Normalization is defined as 10$^{-14}$/4$\pi$D$^2\times \int n_H\,n_e dV$, where $D$ is distance in [cm], n$_\mathrm{H}$ is the hydrogen density [cm$^{-3}$], $n_e$
is the electron density [cm$^{-3}$], and $V$ is the volume [cm$^{3}$]. Fluxes are absorption-corrected and error values are in the 90\% confidence level for every single parameter. Abundances are relative to the solar values of \cite{anders1989}.}
\end{center}
\end{table*}
%\end{landscape}
%%%%%%%%%%%%%%%%%%%%%%%%%%%%%%

%%%%%%%%%%%%%%fig-2%%%%%%%%%%%%%%%
\begin{figure}
\centering
\includegraphics[width=6.6cm]{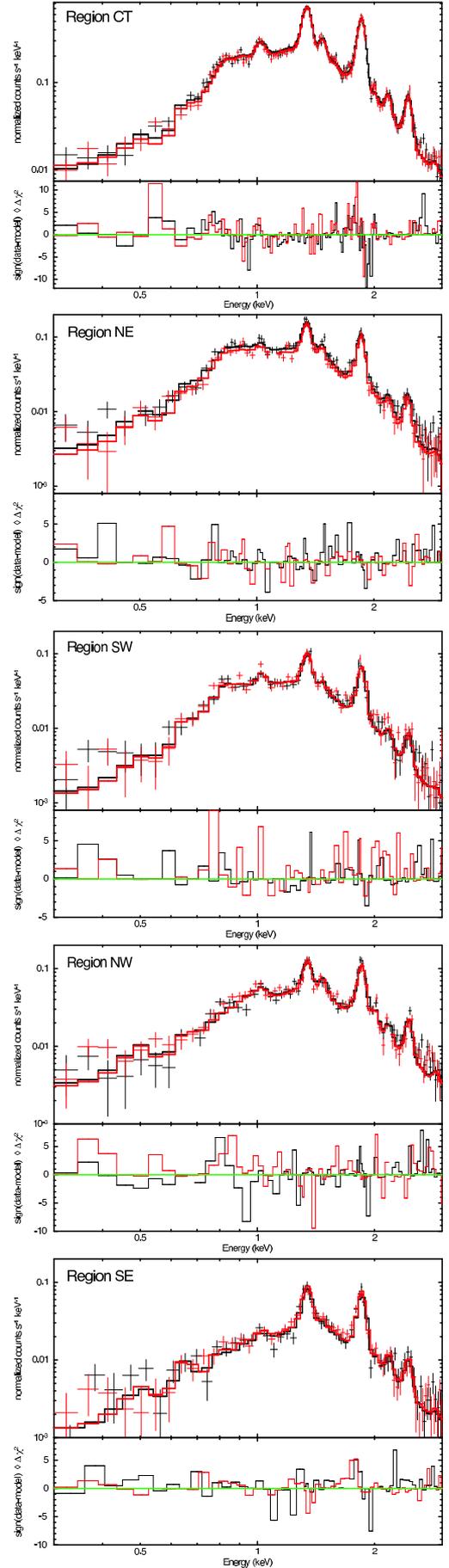}
\caption{{\it XMM-Newton} MOS1/2 spectra of the five SNR selected regions (Figure \ref{fig1}). Solid lines indicate the best-fit VPSHOCK model (see Table~\ref{spectable}). Lower panels are the $\chi^{2}$ fit residuals.}
\label{Fig2}
\end{figure}
%%%%%%%%%%%%%%%%%%%%%%%%%%%%%%%%%%%%%%%%%%%

%%%%%%%%%%%%%%fig%%%%%%%%%%%%%%%
\begin{figure*}
\centering
\includegraphics[width=8cm]{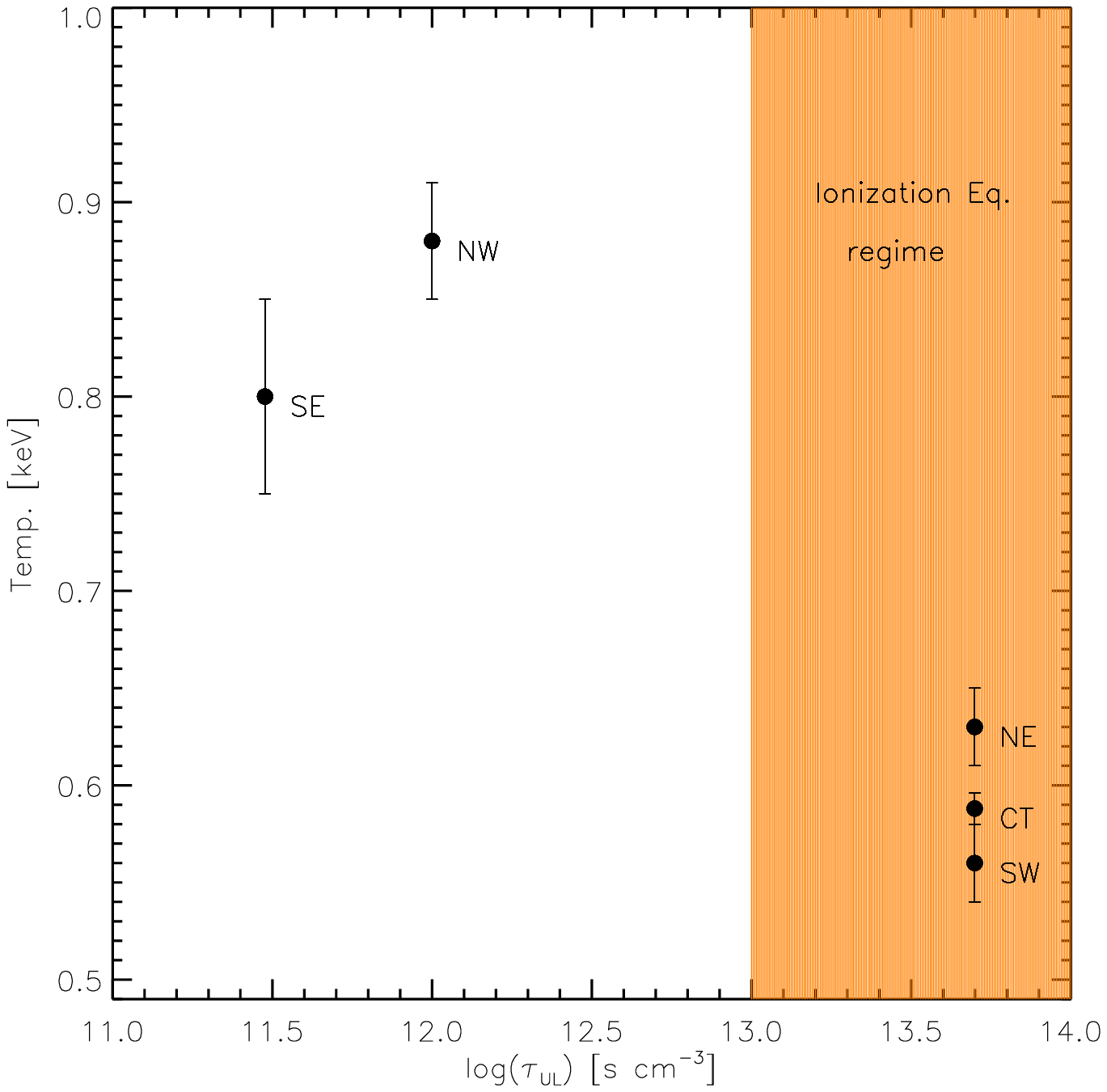}
\includegraphics[width=8cm]{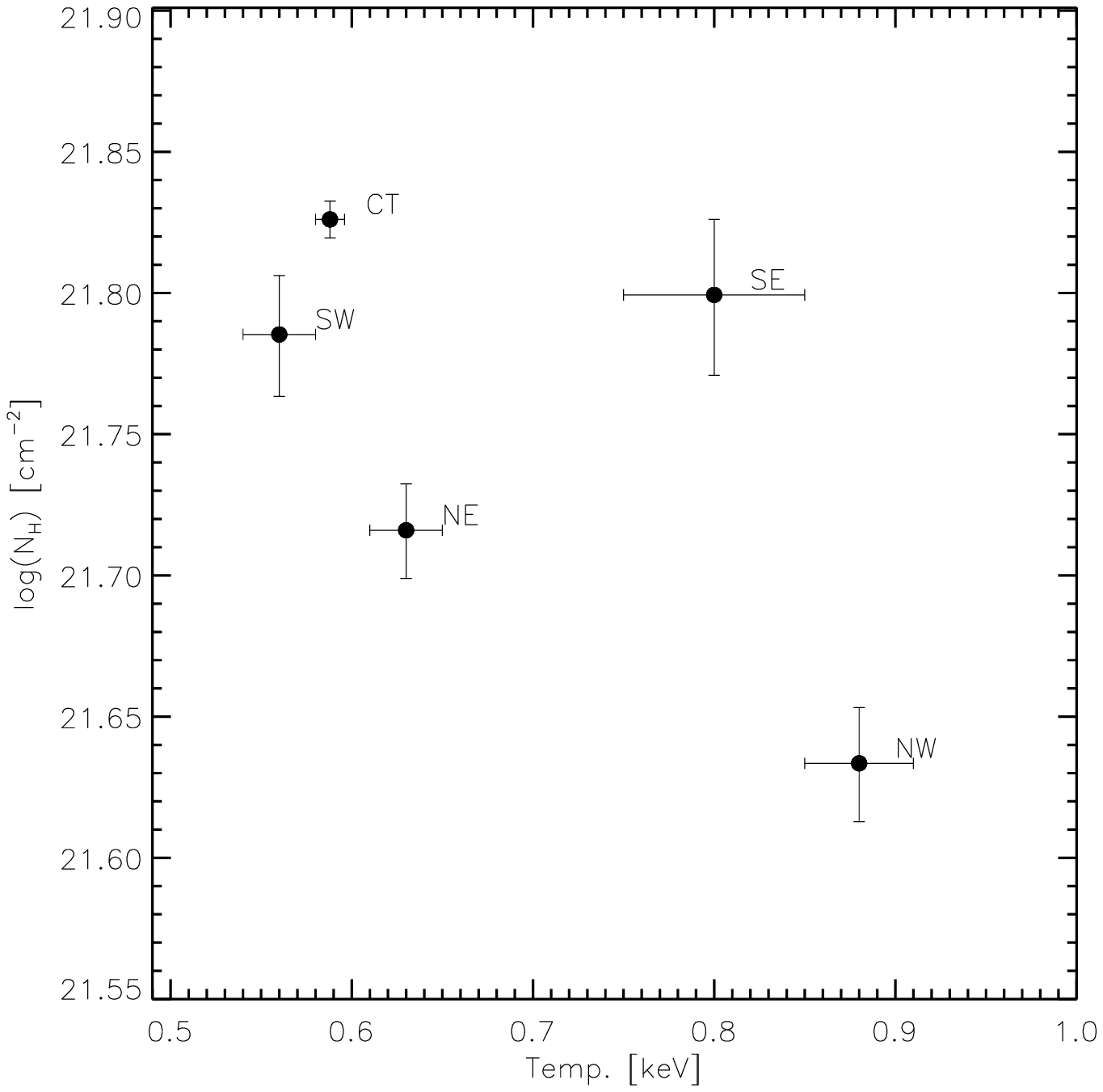}
\caption{{\bf Left Panel:} temperature and ionization timescale ($\tau_\mathrm{ul}$) of the five selected regions. Equilibrium regime ($\tau_\mathrm{ul}>10^{13}$~s~cm$^{-3}$) is highlighted in orange color. {\bf Right Panel:} distribution of the temperature and absorption ($N_\mathrm{H}$) values obtained in the regions considered.}
\label{plots}
\end{figure*}
%%%%%%%%%%%%%%fig%%%%%%%%%%%%%%%

In order to thoroughly study the physical conditions of the plasma, we performed spatially resolved spectral analysis of the SNR using {\it XMM-Newton} data. Based on the morphology observed in the X-ray images, we produced spectra along the North West/South East axis direction in three different regions named: South East (SE), Central (CT) and North West (NW), and for other two zones in the perpendicular line, labeled: South West (SW) and North East (NE) (see Figure~\ref{fig1}, right panel). These regions were chosen large enough to have good photon-statistics and small enough to admit the comparison of their plasma conditions.
Spectra were extracted using {\sc evselect} SAS task with the appropriate parameters for EPIC pn and MOS 1/2 cameras. Ancillary Response Files (ARFs) and Redistribution Matrix Files (RMFs) were produced by use of the {\sc arfgen} and {\sc rmfgen} SAS tasks, respectively. Background was subtracted using the {\it XMM-Newton} Blank Sky files \citep{carter2007} for the same five regions.

Figure~\ref{Fig2} shows the background subtracted spectra of each selected region of the SNR obtained from the {\it XMM-Newton} EPIC MOS1/2 data. In the plots, the spectra are grouped with a minimum of 16~counts per bin. Error bars are quoted at 90$\%$ and $\chi^{2}$ statistic is used. The spectral analysis was performed using the XSPEC package \citep{arnaud1996} working in the 0.3$-$3.0~keV energy range, because no X-ray emission was detected besides this limit. 

The X-ray spectra are dominated by weak and intense emission lines. Following the interactive guide for ATOMDB\footnote{http://cxc.harvard.edu/atomdb/WebGUIDE/index.html} we were able to identify the most prominent lines according to their transition intensities. Observed features in the spectra correspond to atomic transitions of \ion{O}{vii} (0.66~keV), \ion{Ne}{ix} (0.91~keV), \ion{Mg}{xii} (1.47~keV), \ion{Si}{xiii} (1.85~keV), \ion{S}{xv} (2.4~keV) and \ion{Fe}{xxi} (0.99~keV). However, due to the moderate spectral resolution of the EPIC cameras, most features are strongly affected by line blending, biasing the identification and abundance determination of single chemical elements. 

The spectra of the diffuse emission were fitted with various models: a simple Bremsstrahlung model with Gaussian lines, VMEKAL, VNEI, and VPSHOCK, each modified by an absorption interstellar model \citep[PHABS;][]{balucinska1992}. After several tests, we found that the best-fit was computed using a VPSHOCK model, thus confirming the thermal nature of the emitting plasma also found by \cite{slane2002}. 

The X-ray parameters of the best-fit to the diffuse emission spectra in the five regions are given in Table~\ref{spectable}. In Figure \ref{plots} we also present plots to illustrate several important results that we deduce from the spectral analysis. Regions named CT, NE and SW display high values of ionization timescale ($\tau\sim 5\times10^{13}$~s~cm$^{-3}$) suggesting that the plasma has reached the ionization equilibrium, while NW and SE zones show low values ($\tau\la 10^{12}$~s~cm$^{-3}$), indicating that the hot plasma is not yet in equilibrium (see left panel of Figure \ref{plots}). The values of $N_\mathrm{H}$ obtained for each region are similar. However, it is interesting to note that the value of $N_\mathrm{H}$ in the NW region is $30\%$ less than in the opposite SE region. This indicates that the material along the line of sight varies depending on the viewing angle, in accordance to a dense environment in which the remnant expands. Furthermore, the temperatures obtained for the CT, SW and NE regions are significantly lower than the temperatures obtained for the NW and SE zones (see right panel of Figure \ref{plots}). In Table~\ref{spectable}, it can also be observed that different values are obtained for the abundances in CT, NE/SW and NW/SE regions. 

The differences in absorption columns and metal abundances shown here with respect to the previous work by \cite{slane2002} can be explained in the following terms. Firstly, because here we use more recent data from {\it XMM-Newton} that give enough statistics as to produce spatially-resolved high-resolution spectra. Secondly, because of the different spectral model. In their paper, \cite{slane2002} added a power law to the thermal component in order to fit the data in the hard X-rays. This phenomenological power law extends to the soft X-rays, producing a soft excess which must be diminished by a higher absorption column to fit the data in the soft band. However, {\it XMM-Newton} data were well-fitted using only a thermal model, without a power law, and thus giving sistematically lower absorption columns. Moreover, since the weight of the thermal component is enhanced in our fit, due to the absence of a power law, the emission-line profiles are fitted with lower abundances.

The absorption corrected X-ray fluxes computed for the selected regions in the three energy ranges give important information too. While the entire remnant results mostly intense ($\sim$70\% of the total) at soft energies, the rest of the X-ray emission is evenly distributed in the NW and SE regions compared to NE and SW. In the NW and SE regions the ratio hard/medium X-ray contribution results $\sim$0.25, while in the NE and SW zones this ratio becomes $\sim$0.15. So, the emission in the two opposite bright spots is harder than in the rest of the SNR. 

The emission measure (EM) of the five selected zones can bring an estimate of the density and the mass of the corresponding X-ray emitting plasma. Based on the X-ray image, we can roughly assume that the X-ray emission fills the mentioned regions. Located at a distance of 7~kpc, these regions correspond to X-ray emitting volumes of $V_\mathrm{CT} \sim 5.5\times10^{57}$~cm$^{3}$, $V_\mathrm{NW} \sim 2.4\times10^{57}$~cm$^{3}$, $V_\mathrm{SE} \sim 1.8\times 10^{57}$~cm$^{3}$, $V_\mathrm{SW} \sim 0.9\times 10^{57}$~cm$^{3}$ and $V_\mathrm{NE} \sim 2.5\times 10^{57}$~cm$^{3}$. Based on the EM determined by the normalizations obtained from the best-fit model, the electron density of the plasma volumes can be approximated as $n_\mathrm{e}$=$\sqrt{EM/V}$, which gives $n_\mathrm{e} \mathrm{(CT)} \sim n_\mathrm{e} \mathrm{(SW)} \sim 1.1$~cm$^{-3}$, $n_\mathrm{e} \mathrm{(NW)} \sim n_\mathrm{e} \mathrm{(SE)} \sim 1.0$~cm$^{-3}$ and $n_\mathrm{e} \mathrm{(NE)} \sim 0.7$~cm$^{-3}$. The nuclei number density is assumed equal to the electron density. Then, the plasma mass can be estimated as $M = n_\mathrm{e} m_\mathrm{H} V$, where $m_\mathrm{H}$ is the hydrogen-atom mass, resulting: $M_\mathrm{CT} \sim 5.3$M$_{\odot}$, $M_\mathrm{NW} \sim 2.0$M$_{\odot}$, $M_\mathrm{SE} \sim 1.4$M$_{\odot}$, $M_\mathrm{SW} \sim 0.8$M$_{\odot}$ and $M_\mathrm{NE} \sim 1.5$M$_{\odot}$. These are the masses of the ionized plasma, not to be confused with those of the atomic or molecular gas.

Although the physical conditions in the NW and SE regions of G290.1$-$0.8 are different, it is possible to obtain an estimate of the age of the X-ray emitting gas in these regions. The age is determined from the ionization timescale $\tau$, as $t$=$\tau / n_\mathrm{e}$. Assuming this relation, the cooling time of the plasma results in the range of 10$-$30 $\times 10^{3}$~yr. This result agrees with the SNR age deduced by \cite{slane2002}.

\subsection{Mean photon energy map}

From the spatially resolved spectral analysis, we have found that the selected SNR regions present considerably different plasma conditions. However, other areas of the SNR are still in the poor photon-statistics limit, which avoid us from drawing certain conclusions about the energetics of the whole radiation region.

In order to extend the study of this physical anisotropy to the total emission, and to get insight of the thermal structure of the entire SNR, we computed the mean photon energy (MPE) map of the entire SNR. The MPE map is an image where each computed pixel corresponds to the mean energy of the photons detected by MOS CCDs in the 0.3$-$3.0~keV energy band. It provides information about the spatial distribution of the thermal properties of the plasma. Because this characteristic is independent of the spectral model, it can not be corrected by absorption, resulting biased to the hard X-rays. To compute this map, we merged EPIC MOS1 and MOS2 event files and created an image with a bin size of $9\arcsec$ that collects a minimum of 4~counts per pixel everywhere in the remnant. For each pixel, we calculated the mean energy of the photons and then we smoothed the map by using a Gaussian kernel value of $3\sigma$ \citep[see][]{miceli2005}.

\begin{figure}
\centering
\includegraphics[width=8.5cm]{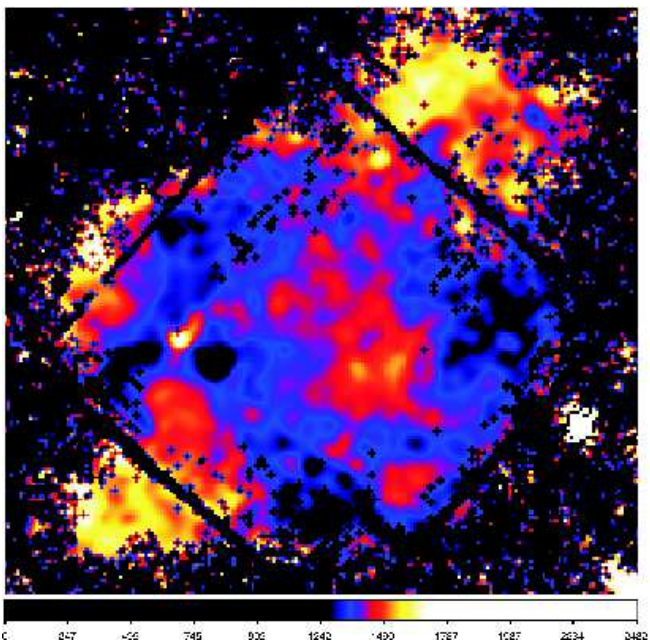}
\caption{MOS mean photon energy map of the 0.3$-$3.0~keV emission (bin size=$9\arcsec$).  Pixels with less than 4 counts have been masked out. The color bar has a linear scale and the color-coded energy range is between 1.2~keV and 1.7~keV.}
\label{mpe}
\end{figure}

The MPE falls in the 1.1$-$1.9~keV range for pixels with less than 10~photons, but converges to a total mean value of $1.43$~keV for the entire SNR emission. According to the smoothed MPE map of G290.1$-$0.8
shown in Figure~\ref{mpe}, we are able to confirm the presence of a higher mean photon energy ($\sim$1.65~keV) in the two opposite bright spots of the remnant, whereas the whole central part has lower mean energies ($\sim$1.3~keV). This result agrees with the spectral analysis showed in the previous subsection where we have shown that the NW and SE zones present higher plasma temperatures and harder X-ray fluxes than the NE, SW and CT regions.

\subsection {Analysis of the cold ISM in the radio wavelengths}

%%%%%%%%%%%%%%fig-5%%%%%%%%%%%%%%%
\begin{figure}
\centering
\includegraphics[width=8.5cm]{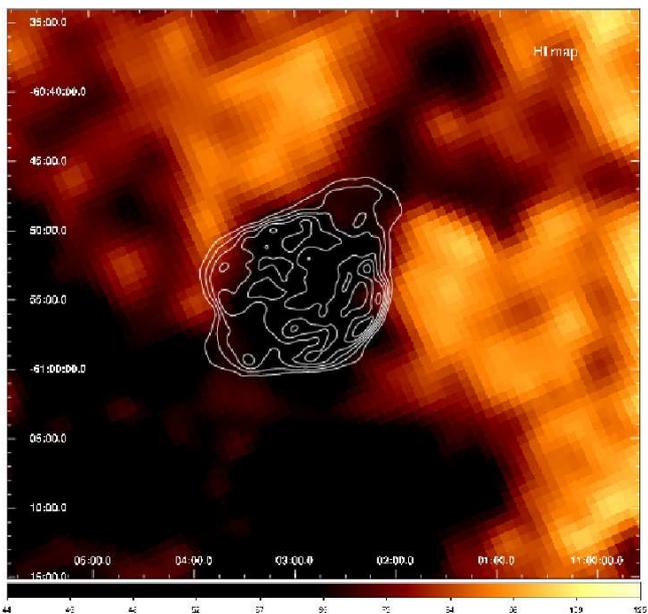}
\caption{\ion{H}{i} map (color-coded) of the ATCA + SGPS combined data covering SNR G290.1$-$0.8 over the
$+5.8$ to $+11.5$~km~s$^{-1}$ velocity range, with the 843 MHz MOST radio contours overlaid. The \ion{H}{i} scale is in units of Kelvin.}
\label{Fig4}
\end{figure}
%%%%%%%%%%%%%%%%%%%%%%%%%%%%%%%%%%%%%%%%%%%5

The \ion{H}{i} line observations studied by \cite{reynoso2006} revealed that the gas distribution around the SNR G290.1$-$0.8 has a complex structure. The SNR lies in a high density ISM, possibly inside a large atomic and molecular cloud. In order to understand if the ISM that surrounds the remnant could affect its evolution, we used the \ion{H}{i} data of the Southern Galactic Plane Survey \citep[SGPS;][]{mccluregriffiths2005} to analyze in more detail the density distribution of the medium. Figure~\ref{Fig4} shows a \ion{H}{i} map over the +5.8 to +11.5~km~s$^{-1}$ velocity range, with the 843 MHz MOST radio continuum contours overlaid. This velocity range is chosen to match the 7~kpc distance to the SNR. In this map, three important features can be observed: first, that the outer radio contour matches very well the \ion{H}{i} diminished emission seen as a rounded central cavity; second, that two clear elongated \ion{H}{i} tube-like structures in the North West edge extend beyond the outer radio boundary; third, that in the South West side a region with apparently higher density seams to be interacting with the SNR edge as previously suggested by \citet{filipovic2005}. If this is a pre-existing structure of the ISM, it might be responsible of the actual radio/X-ray morphology.

The dim emission in the central region can be explained as a result of a real \ion{H}{i} defect due to the passage of the SN shock front or as \ion{H}{i} foreground absorption of the SNR radio continuum. Due to the absence of a detailed model of the opacity of the ISM in the region, no reliable calculations could be drawn to find the \ion{H}{i} density distribution in the SNR interior. However, the highly structured ISM where the SNR is immersed, in correlation with the X-ray morphology observed, favour the scenario of a pre-existing inhomogeneus environment that shaped the observed morphology.

\section {Discussion}

The peculiar morphology observed in G290.1$-$0.8 could be the result of two possible physical scenarios (or a combination of both): i) a spherical explosion in a highly inhomogeneous circumstellar and interstellar medium, or ii) the presence of an undetected central source which produces jet-like structures in two opposite directions. The SNR spectral X-ray analysis, the MPE map, and the information obtained from radio data, can give us clues about the nature of the SNR radio/X-ray morphology.

Our image analysis shows that the remnant has a centrally peaked
morphology, more pronounced in the soft band, thus confirming it as 
belonging to the MM class. Furthermore, our spectral analysis shows
moderate overabundances of \ion{Si}{}everywhere and of \ion{S}{}in the central part,
a situation which is also observed in other MM SNRs \citep[see e.g.][]{bocchino2009}.

Taking into account the characteristic temperatures, ionization timescales and X-ray fluxes obtained in the X-ray spectral analysis, a plausible physical scenario can be outlined. The abrupt drop in the X-ray surface brightness on the North East/South West direction, together with the low temperatures and large ionization timescales deduced from the spectral analysis, indicate that in these extremes the SNR shock front has reached high density clouds where atomic and molecular material has been observed \citep[see][]{filipovic2005}. As a result, the original shock splitted into forward/reverse shocks, moving along opposite directions in both media. While the forward shock is invisible to X-rays, the reverse shock is bright and clearly less energetic than the original shock. The material in the reverse shock cooled down, achieving the equilibrium ionization. Due to the presence of this reverse shock, the central part of the SNR ejecta brightens, as is shown by \cite{zhou2011}. On the contrary, in the North West/South East direction, due to the presence of a pre-existing low density tube-like cavity in the ISM, the primary shock front easily propagated to larger distances preserving its original energy. Thus, in the NW and SE regions the plasma still presents high temperatures and it is in the non-equilibrium ionization regime. In this sense, the radio/X-ray morphology of the emitting plasma and its spectral properties can be explained in terms of the pre-existing ISM density distribution in the surroundings of the SNR.

After fitting the spectra with several models, we found that there is no significant contribution coming from a non-thermal component. As we already metioned, the thermal plasma is at different temperatures and ionization states along the remnant. All the emitting regions present chemical abundances that correspond to ejected material. Moreover, in order to search for a contribution from the ISM to the X-ray emission in the bright central (CT) region, we studied how the spectral fitting was modified by adding a second thermal component with solar abundance to the model. However, a second component did not improve the fit. The data was well-fitted with a single-temperature thermal model, suggesting that no contribution from the ISM is present at all.

\begin{figure}
\centering
\includegraphics[width=8.5cm]{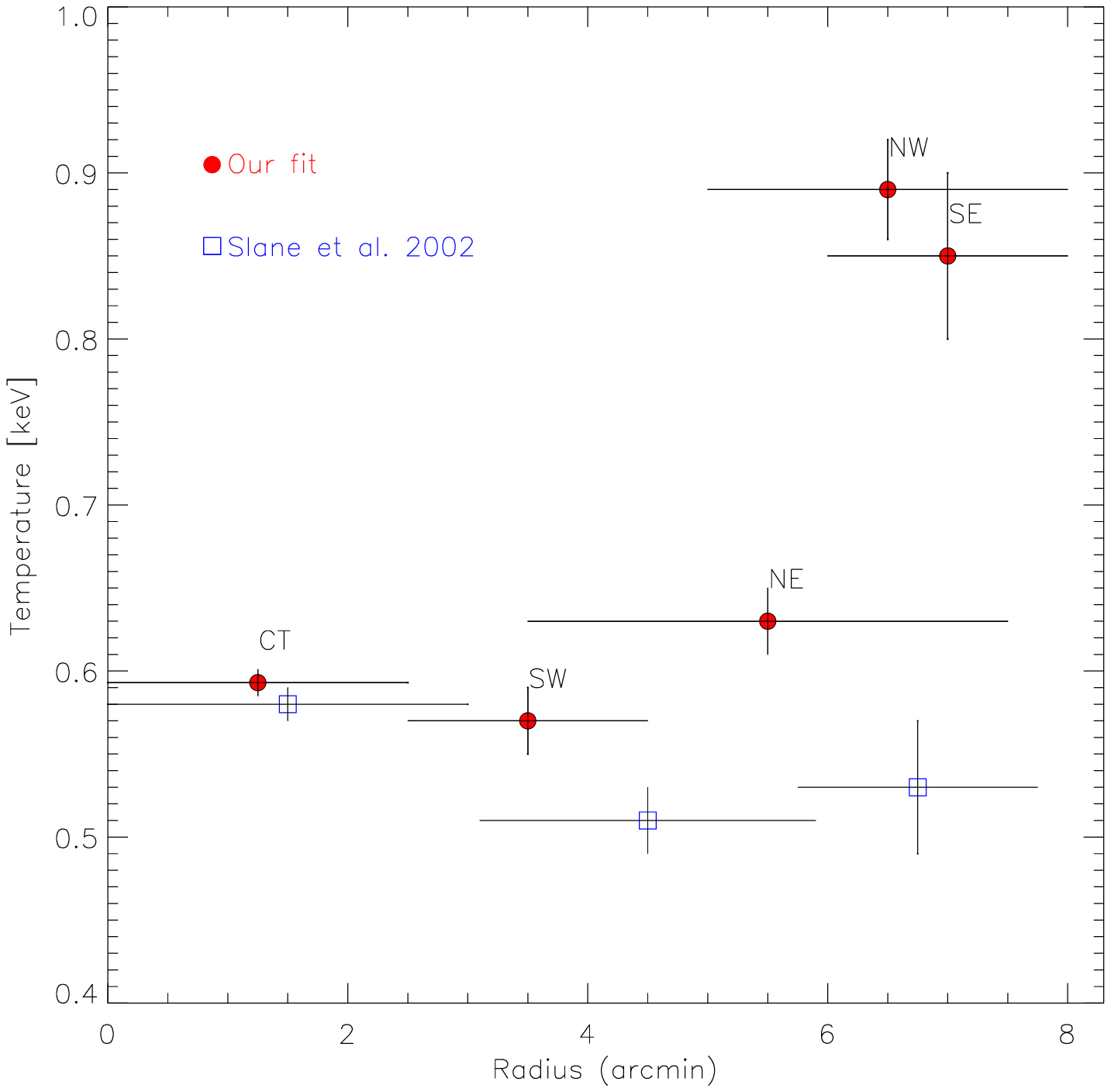}
\includegraphics[width=8.5cm]{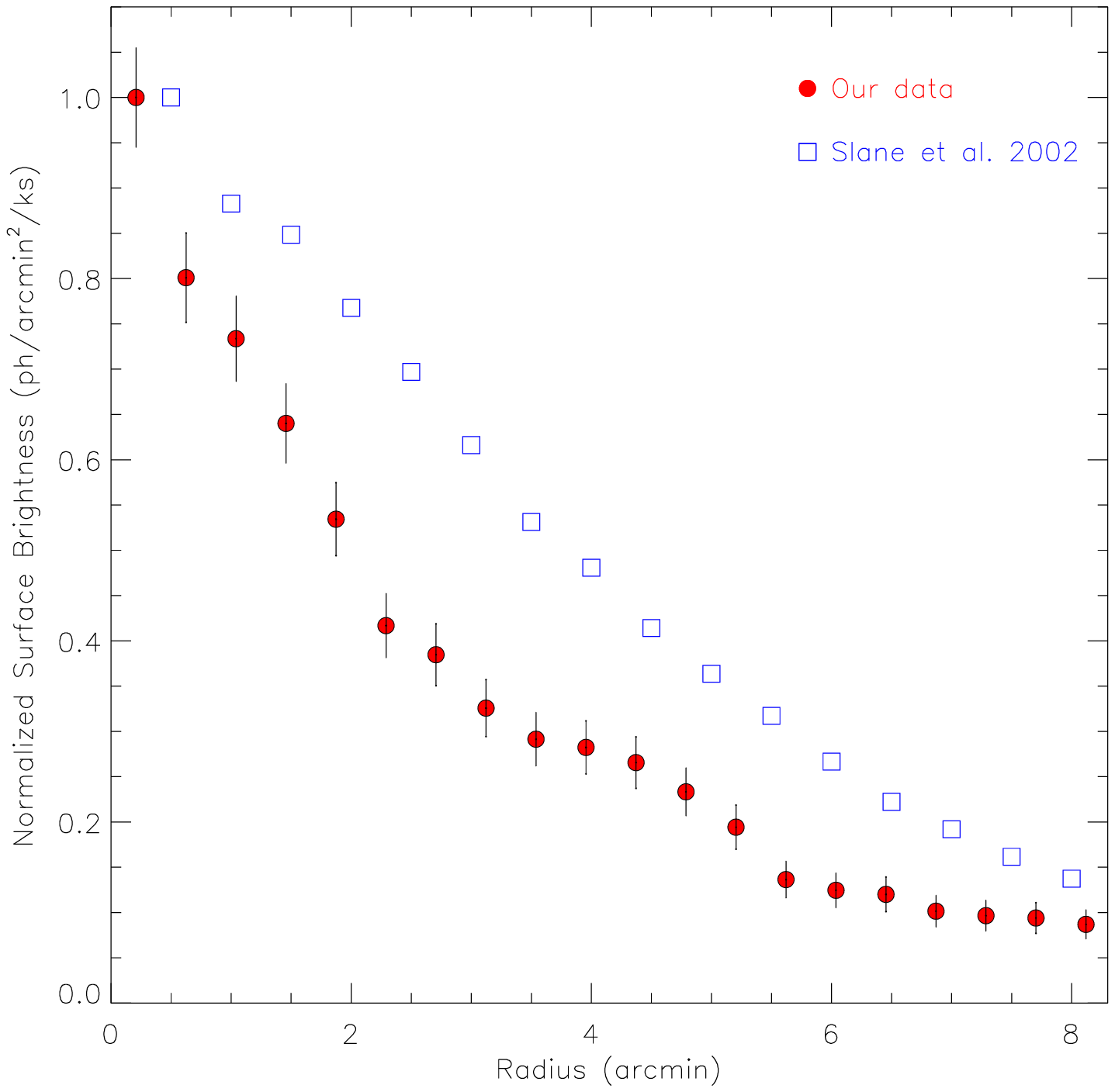}
\caption{{\bf Upper panel:} Comparison of the temperature distribution derived from our {\it XMM-Newton} data analysis with the radial profile presented by \cite{slane2002}. {\bf Lower panel:} same as the upper panel for the surface-brightness profiles.}
\label{Fig6}
\end{figure}

In order to make a quantitative comparison of our results with those obtained by \cite{slane2002} using {\it ASCA} data, in the upper panel of Figure \ref{Fig6}, we present temperature radial profiles of the SNR, derived from our spatially resolved spectral analysis (see Table \ref{spectable}). In addition, in the lower panel, we also show normalized surface-brightness profiles, resulting from the photon counting of MOS1/2 cameras. In both cases, we placed the origin of the radial profiles at the proyected geometrical centre of the SNR. After normalization, both {\it ASCA} and {\it XMM-Newton} surface-brightness profiles show the same trend. However, our data are always below the {\it ASCA} profile. On the contrary, our temperature profile is higher than the profile derived from the {\it ASCA} data all through the remnant. In particular, strong differences occur in the most outer regions, where the ejected material, out of ionization equilibrium, presents temperatures remarkably higher than the rest of the remnant. We suggest that this is a consequence of the boundary conditions imposed by the ambient where the remnant developed, with a pre-existing elongated cavity in the North East-South West direction, and two dense regions in the opposite sides. These conditions that we interpret as fundamental tracers of the observed morphology, are not taken into account in the model proposed by \cite{slane2002}.

%%%%%%%%%%%%%%fig-6%%%%%%%%%%%%%%%
%%%%%%%%%%%%%%%%%%%%%%%%%%%%%%%%%%%%%%%%%%%

Despite that in the analysis performed by \cite{slane2002}, the cloudy
model from \cite{white1991} is the best choice to explain the
observations, they also point out that some problems remain, i.e. they
found longer cloud evaporation time than the age deduced for the
remnant, and the observed X-ray profile results steeper than
expected. Moreover, the cloudy model represents a situation in which a
remnant expands into uniform media. Our analysis shows clearly that this
is not the case. Finally, this model can not explain the presence of any
ejecta in the central region, whereas we have detected ejecta-like
abudances pattern in the centre of this remnant. Therefore, an
explanation for the centrally bright morphology in terms of the result
of shock reflected off dense clouds in the environment, which
additionally compress the ejecta in the centre \citep[see
e.g.][]{zhou2011}, seems much more plausible than the evaporative model
proposed by \cite{slane2002}. 

In order to disentangle this point, a quantitative comparison of
our results with the model proposed by \cite{zhou2011} would be
helpful. However, their simulations were fitted to reproduce the
morphology of SNR W49B, and developing a full numerical
model for G290.1--0.8 is out of the scope of this paper.

The fact that all through the remnant the \ion{Fe}{} abundance is low ($\la$0.1) favours the possibility of a supernova originated as the result of the collapse of a high-mass star. This is also confirmed by a quick comparision between the \ion{Mg}{}/\ion{Si}{}, \ion{S}{}/\ion{Si}{} and \ion{Fe}{}/\ion{Si}{} abudances ratio (0.5, 0.8 and 0.05 respectively, in the centre region), which are broadly consistent with the 25$-$30 M$_{\odot}$ model of \cite{woosley1995} and very inconsistent with any degenerate scenario (see, e.g. table 3 of \cite{troja2008} and references therein). 
Furthermore, a strong bipolar wind from the progenitor star could be the responsible of the elongated structured environment, because high-mass stars can deposit $\sim$10$^{51}$~erg of kinetic energy to their environment through a strong asymmetric wind \citep{lamers1999}.

\section{Conclusions}

We have analyzed radio, X-ray and \ion{H}{i} data of SNR G290.1-0.8 to investigate the origin of the peculiar morphology observed at radio/X-ray frequencies, as well as the interaction of the source with the surrounding ISM. The gathered multiwavelength information supports an astrophysical scenario in which the radio and X-ray morphologies are the result of a spherical explosion in a high density medium presenting a pre-existing  low density cavity with a tube-like structure in the North West/South East direction.

The X-ray spectral analysis and the MPE map confirm that the physical conditions of the plasma are not homogeneous throughout the remnant. While the X-ray emission in the CT, NE and SW regions have reached the ionization equilibrium, in the NW and SE zones the emitting plasma remains in the non-equilibrium ionization regime. Furthermore, the temperatures in the CT, SW and NE parts are significantly lower than those obtained for the NW and SE regions. The chemical abundances are also different, and in all cases correspond to an ejecta component.

Based on the results of this X-ray analysis, we suggest 
that a standard scenario involving one core-collapse supernova origin can be enough to explain the particular morphology here observed, if the action of the pre-existent surrounding environment is taking into account. In this case, a strong bipolar wind from the progenitor star could be responsible of the tube-like elongated cavities observed at the radio wavelengths. 

Although the evidence here obtained supports that the SNR is the result of a spherical explosion in an anisotropic high density medium, the possibility that an undetected point-like object inside the SNR, as a neutron star, be contributing to the origin of the X-ray morphology, cannot be ruled out.

Deeper {\it Chandra} observations could help to reveal more details of the X-ray emission and to discern the presence of a compact point-like X-ray source within the remnant.

\begin{acknowledgements}
      
We are grateful to the referee for his valuable suggestions and comments which helped to improve the paper. The authors acknowledge support by DGI of the Spanish Ministerio de Educaci\'on y Ciencia under grants AYA2010-21782-C03-03, FEDER funds, Plan Andaluz de Investigaci\'on Desarrollo e Innovaci\'on (PAIDI) of Junta de Andaluc\'{\i}a as research group FQM-322 and the excellence fund FQM-5418. G.E.R, J.A.C. and J.F.A.C. are researchers of CONICET. J.F.A.C was supported by grant PIP 114-2011-0100285 (CONICET), J.A.C by grant PICT 2008-0627, from ANPCyT and PIP 2010-0078 (CONICET) and G.E.R. by grant PICT 07-00848 BID 1728/OC-AR (ANPCyT) and PIP 2010-0078 (CONICET). J.L.S. and G.E.R. acknowledges support by the Spanish Ministerio de Innovaci\'on y Tecnolog\'ia under grant AYA2008-06423-C03-03 and AYA2010-21782-C03-01. F.B. acknowledges partial support from ASI-INAF agreement n. I/009/10/0.

\end{acknowledgements}

\end{document}